\documentclass[11pt,a4paper]{article}
\usepackage[latin1]{inputenc}
\usepackage{amsmath}
\usepackage{amsfonts}
\usepackage{amssymb}
\usepackage{graphicx}
\DeclareGraphicsExtensions{.pdf}
\author{Sumit Ghosh}
\title{GUP and Correction to BTH}
\begin{document}

\title{{\bf{\Large Generalised Uncertainty Principle, Remnant Mass \\ and \\ Singularity Problem in Black Hole Thermodynamics}}}

\author{{\bf {\normalsize Rabin Banerjee}$
$\thanks{E-mail: rabin@bose.res.in}},\,
{\bf {\normalsize Sumit Ghosh}$
$\thanks{E-mail: sumit.ghosh@bose.res.in}}\\
{\normalsize S.~N.~Bose National Centre for Basic Sciences,}
\\{\normalsize JD Block, Sector III, Salt Lake, Kolkata-700098, India}
\\[0.3cm]}

\maketitle
\begin{abstract}
We have derived a new Generalised Uncertainty Principle (GUP) based on certain general assumptions. This GUP is consistent with predictions from string theory. It is then used to study Schwarzschild black hole thermodynamics. Corrections to the mass-temperature relation, area law and heat capacity are obtained. We find that the evaporation process stops at a particular mass, referred as the remnant mass. This is instrumental in bypassing the well known singularity problem that occurs in a semiclassical approach.
\end{abstract} 

\section{\bf{Introduction}}
The introduction of gravity into quantum field theory brings an observer independent minimum length scale in the picture \cite{aml}. A minimal length also occurs in string theory \cite{amt}, non commutative geometry \cite{grl} or can be obtained from gedanken experiment \cite{scr}. This minimal length is expected to be close or equal to Planck length ($L_P$). The manifestation of the inclusion of a minimal length in theories has been observed from different perspectives - the generalised uncertainty principle (GUP), modified dispersion relation (MDR), deformed special relativity (DSR), to name a few. The modification or deformation affect the well known semiclassical laws of black hole thermodynamics \cite{arz,vag,nuc,bin}. For instance, the black hole entropy is no longer proportional to the horizon area \cite{hwk,bkn}. 
Another interesting result is the existence of a remnant mass of a black hole. The existence of a remnant mass of a black hole is verified by different approaches - using a generalised uncertainty principle \cite{adl} or analysing the tunnelling probability \cite{xng}.   
  
In this paper we study the modifications to the laws of Schwarzschild black hole thermodynamics by starting from a new GUP which is derived from certain basic assumptions. The consequences of these modifications are investigated in details. We show the existence of a critical (ans also singular) mass for the black hole below (at) which the thermodynamic entities become complex (ill defined). However both the critical and singular mass are less than another mass - the remnant mass. Our analysis reveals that the black hole evaporation does not lead to a singularity. This process terminates at a finite mass which we call the remnant mass. Since, as already stated, both the critical and singular masses are less than the remnant mass, the problematic situations are avoided.\\

Section 2 gives the derivation of the GUP. Corrections induced by this GUP to the black hole thermodynamics are given in section 3. The connection of remnant mass with the singularity problem is discussed in section 4. Section 5 contains the discussions.   
\section{\bf{A Generalised Uncertainty Principle (GUP)}}

A particle with energy close to the Planck energy $E_{P}$ will disturb the space time significantly at least upto a length of the order of the Planck length. It is very natural to take the metric to be a function of the particle energy\footnote{This is the effect of back reaction \cite{lou}.} \cite{hosd} . One can find the explicit dependence by solving corresponding Einstein equation where the right hand side is given by the energy momentum tensor of the particle. If we assume that the a particle field is a linear superposition of plane wave solutions ($ \thicksim e^{ik^\mu x_\mu}$), then one can easily guess that on quantisation the particle momentum$(p)$ and energy$(E)$ may be non linear in wave vector$(k)$ and angular frequency$(\omega)$ \cite{hosd,jo} . In general we may write
\begin{eqnarray}
\it{k^\mu = f(p^\mu)} 
\label{f}
\end{eqnarray} 

It is easy to show that both $k^\mu$ and $p^\mu$ can transform like a Lorentz vector only for special types of function $f$. The standard form is the obvious $p^\mu = \hbar k^\mu$ and a more general form is $p^\mu=\phi \left(k^\nu k_\nu \right)k^\mu$ where $\phi \left(k^\nu k_\nu\right)$ is a scalar function of the invariant $\left(k^\nu k_\nu\right)$; the more general form is clearly equivalent to generalising Planck's constant to a function. For simplicity in this paper we will forego Lorentz invariance and consider the following relations \cite{hosc2},

\begin{eqnarray}
\it{k = f(p)   \hspace{1cm}   \omega = f(E)} 
\label{kp}
\end{eqnarray} 
 
The function $f$ satisfies certain properties \cite{hosd,hosc} :
\begin{enumerate}
\item The function $\left(f\right)$ and its inverse $\left(f^{-1}\right)$ have to be an odd function to preserve parity.
\item For small momentum/energy ($E<<E_{P}$) the function should be chosen to satisfy the relationship $p=\hbar k$ . 
\item We assume the existence of a minimum length, identified as the Planck length $(L_P)$ \cite{aml,hosd,hosc} that cannot be resolved. So the wave vector $k=f(p)$ should have an upper bound $\dfrac{2\pi}{L_P}$. Since the wave vector $k=f(p)$ shows a saturation with respect to the momentum $p$, the momentum $p = f^{-1}(k)$ will be a monotonically increasing function of $k$.    
\end{enumerate}

We also assume that the commutation relations 
\begin{eqnarray}
[ x , k ] = i   \;\;\;,\;\;\; [x,p(k)]= i\dfrac{\partial p}{\partial k}
\label{cmu}
\end{eqnarray}

hold which lead to a uncertainty relation \cite{book}
\begin{eqnarray}
\Delta p \Delta x \geqslant \vert \left\langle \dfrac{1}{2}[x,p] \right\rangle \vert \; =\; \dfrac{1}{2}\vert \left\langle \dfrac{\partial p}{\partial k} \right\rangle \vert
\label{up}
\end{eqnarray} 

Observe that we are not using the field theory commutator between the field and its conjugate momentum. Rather our analysis is based on the algebra ({\ref{cmu}}) which is plausible.

The properties of the function $f(p)$, enlisted below ({\ref{kp}}) cannot be satisfied by a finite order polynomial. A possible choice is
\begin{equation}
k = f(p) = \dfrac{1}{L_P} \sum_{i=0} ^\infty a_i (-1)^i {\left(\dfrac{L_P \;p}{\hbar} \right)}^{2i+1}
\label{k}
\end{equation}

Only odd powers of $p$ appear in the polynomial ensuring that $f(p)$ is odd in $p$ (property 1). The coefficients  $\lbrace a_i \rbrace $  are all positive with $a_0 = 1$ (to satisfy $p=\hbar k$ at small energy(property 2)). The factor $(-1)^i$ ensures saturation (property 3). The third property further implies that for $p\rightarrow \infty ,\hspace{.5cm} k\rightarrow \dfrac{2\pi}{L_P}$, i.e.

\begin{eqnarray}
\sum_{i=0} ^\infty a_i (-1)^i {\left(\dfrac{L_P\;p}{\hbar}\right)}^{2i+1} & \longrightarrow\; 2\pi  
\end{eqnarray}

From ({\ref{k}}) we get
\begin{equation}
\dfrac{\partial k}{\partial p} = \dfrac{1}{\hbar} \sum_{i=0} ^\infty a_i (2i+1) (-1)^i \left(\dfrac{pL_P}{\hbar} \right)^{2i}
\end{equation}
Inverting this we obtain
\begin{equation}
\dfrac{\partial p}{\partial k} = \hbar \sum_{i=0} ^\infty a'_i \;\left(\dfrac{pL_P}{\hbar}\right)^{2i}
\end{equation}
where the new coefficients of expansions $\lbrace a'_i\rbrace$ are functions of $\lbrace a_i\rbrace$. It is very easy to show that the first coefficient of this inverted series will be inverse of $a_0$, i.e. 1. 

Hence the GUP following from ({\ref{up}}) takes the form,
\begin{equation}
\Delta x \Delta p \geqslant \left\langle  \dfrac{\hbar}{2} \sum_{i=0} ^\infty a'_i \left(\dfrac{L_P p}{\hbar}\right)^{2i} \right\rangle \geqslant \dfrac{\hbar}{2} \sum_{i=0} ^\infty a'_i \left(\dfrac{L_P}{\hbar}\right)^{2i} \left((\Delta p)^2 + \left\langle p \right\rangle ^2 \right)^i
\end{equation}
where we have used $\left\langle p^{2i} \right\rangle  \geqslant \left\langle p^{2} \right\rangle^i$. For minimum position uncertainty we put $\left\langle p \right\rangle=0$ and our GUP becomes 

\begin{equation}
\Delta x \Delta p \geqslant \dfrac{\hbar}{2}\sum_{i=0} ^\infty a'_i \left(\dfrac{L_P \Delta p}{\hbar}\right)^{2i}
\label{gup}
\end{equation}

Note that all $a'_i$'s are positive. If only the first two terms in ({\ref{gup}}) are considered we reproduce the GUP predicted by string theory \cite{scr,mag}.\\

\section{{\bf Thermodynamics of  Schwarzschild black hole with corrections}}
The object of this section is to use the GUP ({\ref{gup}}) to evaluate different thermodynamic entities of a Schwarzschild black hole and thereby find relations among them.

Let us consider a Schwarzschild black hole with mass M. Let a pair (particle-antiparticle) production occur near the event horizon. For simplicity we consider the particles to be massless\footnote{For massive particle the expression for temperature ({\ref{p}}) will be modified.}. The particle with negative energy falls inside the horizon and that with positive energy escapes outside the horizon and observed by some observer at infinity. The momentum of the emitted particle$(p)$, which also characterises its temperature $(T)$ \footnote{For simplicity we consider the emitted spectrum to be thermal.}, is of the order of its momentum uncertainty $(\Delta p)$ \cite{adl}. Consequently
\begin{eqnarray}
T =\dfrac{\Delta pc}{k_B}
\label{p} 
\end{eqnarray}
For thermodynamic equilibrium, the temperature of the particle gets identified with  the temperature of the black hole itself.\\

It is now possible to relate this temperature with the mass (M) of the black hole by recasting the GUP ({\ref{gup}}) in terms of T and M. In that case the GUP has to be saturated
\begin{equation}
\Delta p \Delta x = {\epsilon}_1 \dfrac{\hbar}{2} \sum_{i=0} ^\infty a'_i \;\left(\dfrac{\Delta pL_P}{\hbar}\right)^{2i} 
\end{equation} 
where the new dimensionless parameter ${\epsilon}_1 $ is a scale factor  saturating the uncertainty relation. We can later adjust it by calibrating with some known result. We add that the product of $\Delta x$ and $\Delta p$ may be arbitrarily large but we assume that the lower limit can be achieved. \\

Near the horizon of a black hole the position uncertainty of a particle will be of the order of the Schwarzschild radius of the black hole \cite{vag,adl}, 
\begin{eqnarray}
\Delta x \; = \; {\epsilon}_2 \dfrac{2GM}{c^2}
\label{x}
\end{eqnarray}
The new dimensionless parameter ${\epsilon}_2 $ is  introduced as a scale factor and will be calibrated soon. 
\\

Substituting the values of $\Delta p$ ({\ref{p}}) and $\Delta x$ ({\ref{x}}) in ({\ref{gup}}), the GUP is recast as
\begin{eqnarray}
 M \;=\; \epsilon \dfrac{M_P}{4}\sum_{i=0} ^\infty a'_i \;{\left( \dfrac{k_B T}{M_P c^2}\right)}^{2i-1}
 \label{m1} 
\end{eqnarray}
(where we have used the relations $\epsilon=\dfrac{{\epsilon}_1 }{{\epsilon}_2},~  M_P =\dfrac{L_P c^2}{G}$ and $\dfrac{c\hbar}{L_P}=M_P c^2 ~~,~M_P$ being the Planck mass.)\\
 
In the absence of correction due to quantum gravity effects only $a'_0 = 1$ will survive and we should reproduce the semi classical result. In this approximation ({\ref{m1}}) reduces to
\begin{equation}
M = \epsilon \dfrac{M_P^2c^2}{4 k_B T}
\end{equation}
 
 This will fix $\epsilon$. Comparison with the standard semi classical Hawking temperature \cite{hwk} $\left( T_H = \dfrac{M_P^2c^2}{8\pi Mk_B}\right)$ yields $\epsilon =\dfrac{1}{2\pi}$.\\
  
So the mass temperature relationship is
\begin{eqnarray}
 M \;=\;  \dfrac{M_P}{8\pi}\sum_{i=0} ^\infty a'_i \;{\left( \dfrac{k_B T}{M_P c^2}\right)}^{2i-1} 
\label{m}
\end{eqnarray} 
\\

The heat capacity of the black hole , by definition, is given by
\begin{eqnarray}
C \;=\; c^2 \dfrac{dM}{dT}
\label{cd}
\end{eqnarray}

Therefore from ({\ref{m}}) we find that
\begin{eqnarray}
C \; =\;\dfrac{k_B}{8\pi}\sum_{i=0} ^\infty a'_i\;(2i-1) \;\left(\dfrac{k_B T}{M_P\; c^2}\right)^{2i-2}
\label{c}
\end{eqnarray} 

The nature of the heat capacity will become more illuminating if we express it in terms of particle energy $E=k_B T$ and Planck energy $E_P=M_P c^2$. Then
\begin{equation}
C \; =\;\dfrac{k_B}{8\pi}\sum_{i=0} ^\infty a'_i\;(2i-1) \;\left(\dfrac{E}{E_P}\right)^{2i-2}
\label{ce}
\end{equation}

For $E<<E_P$ the first term will predominate, and since it is with a negative signature the heat capacity will also be negative in this region. The heat capacity increases monotonically as $E\rightarrow E_P$. There will be a point at which the heat capacity vanishes. We consider the corresponding temperature to be the maximum temperature attainable by a black hole during evaporation. The process stops thereafter. 
  
So a Schwarzschild black hole with a finite mass and temperature, by radiation process, loses its mass and in turn its temperature increases. This state corresponds to a negative heat capacity. Then it attains a temperature at which $\dfrac{dM}{dT}$ becomes zero (zero heat capacity), i.e there will be no further change of black hole mass with its temperature. The radiation process ends here with a finite remnant mass with a finite temperature.\\

One can also determine the black hole entropy(S) in a similar way. According to the first law of black hole thermodynamics it is given by 
\begin{equation}
S \;=\; \int\dfrac{c^2dM}{T} 
\end{equation}
For technical simplification this definition is expressed in terms of the heat capacity ({\ref{cd}}). Then exploiting ({\ref{c} }) and carrying out the above integration we finally obtain
\begin{eqnarray}
S &=& \int\dfrac{CdT}{T}\nonumber\\
  &=& \dfrac{k_B}{16\pi} \left[  \left( \dfrac{M_Pc^2}{k_BT} \right)^2  \;+a'_1 \ln \left(\dfrac{k_B T}{M_Pc^2} \right)^2 \; + \sum_{i=2} ^\infty a'_i\;\dfrac{(2i-1)}{(i-1)} \; \left(\dfrac{k_B T}{M_Pc^2} \right) ^{2(i-1)} \right]
\label{s}
\end{eqnarray} 

If we want to express the heat capacity and the entropy in terms of the mass we have to obtain an expression for $T^2$ in terms of M. We can do this by squaring ({\ref{m}})
\begin{eqnarray}
\left(\dfrac{8\pi M}{M_P} \right)^2 &=& \left( \dfrac{M_Pc^2}{k_BT} \right)^2 + 2a'_1 + \left( {a'_1} ^2 + 2 a'_2 \right)  \left( \dfrac{k_BT}{M_Pc^2} \right)^2 + 2\left( a'_1 a'_2 +  a'_3 \right)  \left( \dfrac{k_BT}{M_Pc^2} \right)^4 +\ldots\nonumber\\
\label{mt}
\end{eqnarray}
Then considering a finite number of terms, dictated by the order of the approximation, we can obtain an expression for $\left( \dfrac{k_BT}{M_Pc^2} \right)^2$ in terms of M by inverting ({\ref{mt}} ).

\subsection{{\bf First order correction}}
We will next discuss the effect of first order correction. Consequently we can neglect the contribution of $\left( \dfrac{k_BT}{M_Pc^2} \right)^2$ and higher order terms in ({\ref{mt}}). Then we get 
\begin{equation}
\left( \dfrac{k_BT}{M_Pc^2} \right)^2\;=\;\dfrac{1}{\left(\dfrac{8\pi M}{M_P}\right)^2 - 2a'_1}
\label{mt1}
\end{equation}

The critical mass below which the temperature becomes a complex quantity is given by 
\begin{equation}
M_{cr}=\dfrac{\sqrt{2a'_1 }}{8\pi}M_P
\label{mcr1}
\end{equation}
We will soon show that the evaporation process terminates with a mass greater than this.\\

Substituting ({\ref{mt1} }) in ({\ref{s}}) we get
\begin{eqnarray}
\dfrac{S}{k_B} &=& \dfrac{S_{BH}}{k_B}-\dfrac{2a'_1}{16\pi}-\dfrac{a'_1}{16\pi}ln\left(\dfrac{S_{BH}}{k_B}-\dfrac{2a'_1}{16\pi}\right)-\dfrac{a'_1}{16\pi}ln(16\pi)
\label{s1}
\end{eqnarray} 
where $S_{BH}=k_B\dfrac{4\pi M^2}{M_P^2}$ is the Bekenstein-Hawking entropy. 

We now obtain the area theorem from equation ({\ref{s1}}). This theorem will appear more tractable if we introduce a new variable $A'$ (reduced area) defined as
\begin{eqnarray}
A'\;=\;16\pi \dfrac{G^2M^2}{c^4}-\dfrac{2a'_1}{4\pi} \dfrac{G^2M_P^2}{c^4}\;=\;A-\dfrac{2a'_1}{4\pi}L_P^2 
\label{a1}
\end{eqnarray}
where A is the usual area of the horizon.

In terms of the reduced area the expression for entropy takes a familiar form
\begin{eqnarray}
\dfrac{S}{k_B} &=& \dfrac{A'}{4L_P^2} - \dfrac{a'_1}{16\pi} \ln \left(\dfrac{A'}{4L_P^2}\right) -\dfrac{a'_1}{16\pi}\ln(16\pi)
\label{sa1}
\end{eqnarray}
This is the area theorem in presence of the GUP ({\ref{gup}}) upto the first order correction. The usual Bekenstein-Hawking semiclassical area law is reproduced for $a'_1=0$. \\

The important feature of ({\ref{sa1}}) is that entropy is explicitly expressed as a function of the reduced area and not the actual area \cite{arz,vag}. It has a singularity at zero reduced area which corresponds to a singular mass given by
\begin{equation}
M_{sing}=\dfrac{\sqrt{2a'_1 }}{8\pi}M_P
\label{ms1}
\end{equation}
 We will subsequently prove that the reduced area is always positive in presence of quantum gravity effect and the singularity is thereby avoided during the evaporation process. 
Observe that, to this order, the critical mass ({\ref{mcr1}}) and the singular mass ({\ref{ms1}}) are identical.  

The variation of temperature and entropy with mass for different values of $a'_1$ is shown in $fig~1,2$. The interesting fact about these curves is their termination at some finite mass for non zero $a'_1$. We will explain the reason in the next section.
\subsection{\bf Second order correction}
In this subsection the various thermodynamic variables are computed upto second order. This implies that terms upto $\left( \dfrac{k_B T}{M_P c^2} \right)^2$ in ({\ref{mt}}) are retained. With a simple rearrangement one can easily obtain 
\begin{eqnarray}
\left(\dfrac{k_BT}{M_Pc^2}\right)^2 &=& \dfrac{ \left[\left(\dfrac{8\pi M}{M_P}\right)^2 - 2a'_1)\right]\pm \sqrt{\left[\left(\dfrac{8\pi M}{M_P}\right)^2 - 2a'_1)\right]^2 - 4({a'_1}^2+2a'_2)} }{2\left({a'_1}^2 + 2a'_2\right)}
\label{mt2}
\end{eqnarray}
Only the (-) sign is acceptable from the $(\pm)$ part, because the (+) sign will not produce the semi classical result if we put $a'_1 = a'_2 = 0$. 

At a first glance, one observes that the first order expression for temperature ({\ref{mt1}}) cannot be retrieved simply  by putting $a'_2=0$. Instead, one has to put ${a'_1}^2+2a'_2=0$, because both ${a'_1}^2$ and $2a'_2$ bear the signature of the second order approximation (see~({\ref{mt}})). To make this retrieval simpler we will rearrange ({\ref{mt2}}) using binomial expansion. 
\begin{eqnarray}
\left(\dfrac{k_BT}{M_Pc^2}\right)^2 &=&\dfrac{1}{\left(\dfrac{8\pi M}{M_P}\right)^2 - 2a'_1} \left[ 1+ \dfrac{\left({a'_1}^2 + 2a'_2\right)}{\left[\left(\dfrac{8\pi M}{M_P}\right)^2 - 2a'_1 \right]^2} +\ldots \right] \nonumber \\
\label{mt2e}
\end{eqnarray}

From this expression it is clear that the first order expression for temperature can be obtained from the second order expression by putting ${a'_1}^2+2a'_2=0$. The critical mass is found by equating the discriminant of ({\ref{mt2}}) to zero. It is given by
\begin{eqnarray}
\left(\dfrac{8\pi M_{cr}}{M_P}\right)^2 = 2a'_1 + 2 \sqrt{{a'_1}^2+2a'_2}
\label{mcr2}
\end{eqnarray}
which is greater than ({\ref{mcr1} }) with first order correction (since $a'_1,a'_2$ are all positive).\\

We conclude this section by computing the corrections (upto second order) to the entropy and the area law. The corrected entropy follows from ({\ref{s}})
\begin{eqnarray}
S&=& \dfrac{k_B}{16\pi} \left[  \left( \dfrac{M_Pc^2}{k_BT} \right)^2  \;+a'_1 \ln \left(\dfrac{k_B T}{M_Pc^2} \right)^2 \; + 3a'_2 \left(\dfrac{k_B T}{M_Pc^2} \right)^2 \right]
\label{st2}
\end{eqnarray} 

Substituting ({\ref{mt2e}}) in ({\ref{st2}}) we get
\begin{eqnarray}
\dfrac{S}{k_B}&=&\left(\dfrac{S_{BH}}{k_B}-\dfrac{2a'_1}{16\pi}\right)-\dfrac{a'_1}{16\pi}ln\left(\dfrac{S_{BH}}{k_B}-\dfrac{2a'_1}{16\pi}\right)+\sum_{j=0}^{\infty } c_j(a'_1,a'_2)\left(\dfrac{S_{BH}}{k_B}-\dfrac{2a'_1}{16\pi}\right)^{-j}-\dfrac{a'_1}{16\pi}\ln(16\pi)\nonumber\\
\label{s2}
\end{eqnarray}
where $S_{BH}=k_B\dfrac{4\pi M^2}{M_P^2}$ is the Bekenstein-Hawking entropy and these new coefficients of expansion $c_j$ are functions of $a'_1$ and $a'_2$. The coefficients $c_j$ will have an explicit dependence on ${a'_1}^2 + 2a'_2$, which follows from ({\ref{mt2e}}). Hence we can get our first order result ({\ref{s1}}) by putting ${a'_1}^2 + 2a'_2=0$.\\

We can now obtain our area theorem in terms of the reduced area ({\ref{a1}}) from ({\ref{s2}}),
\begin{equation}
\dfrac{S}{k_B}\;=\;\dfrac{A'}{4L_P^2} - \dfrac{a'_1}{16\pi} \ln \left(\dfrac{A'}{4L_P^2}\right)+\sum_{j=0}^{\infty } c_j(a'_1,a_2)\left(\dfrac{A'}{4L_P^2}\right)^{-j} -\dfrac{a'_1}{16\pi}\ln(16\pi)
\label{sa2}
\end{equation}
This is the area theorem with second order correction. The expression looks like the standard corrected area theorem \cite{arz,vag}, with the role of the actual area ($A$) being played by the reduced area ($A'$).

Some comments are in order. The singularity is again at zero reduced area, corresponding mass being given by ({\ref{ms1}}). As shown in the next section, this singular mass is also less than the remnant mass with second order correction.


\section{{\bf Remnant mass and singularity problem}}
As discussed earlier here we show that the black hole evaporation terminates at a finite mass which is greater than the either the critical mass $M_{cr}$({\ref{mcr1}},{\ref{mcr2}}) or the singular mass ({\ref{ms1}}). This demonstrates the internal consistency of our calculation scheme. Consequently the usual singularity problem whereby the temperature blows up, is avoided.

\subsection{\bf First order correction} 

Considering the first two terms in the series expansion for the heat capacity ({\ref{c}}), we obtain
\begin{equation}
C\;=\;\dfrac{k_B}{8 \pi} \left[ -\left( \dfrac{M_P c^2}{k_B T} \right)^2  + a'_1 \right]
\end{equation}
Substituting the value of $T^2$ from ({\ref{mt1}})
\begin{equation}
C\;=\;\dfrac{k_B}{8 \pi} \left[ -\left(\left( \dfrac{8 \pi M}{M_P} \right)^2 -2 a'_1\right) + a'_1 \right]
\end{equation}

The variation of heat capacity with mass for different values of $a'_1$ is shown in $fig~3$.\\
												
The collapse of the black hole is terminated when the heat capacity becomes zero. The mass of the black hole now remains unchanged. This mass is called the remnant mass. Its value is obtained by solving 
\begin{equation}
\dfrac{k_B}{8 \pi} \left[ -\left(\left( \dfrac{8 \pi M}{M_P} \right)^2 -2 a'_1\right) + a'_1 \right]\;=\;0
\end{equation}
leading to,
\begin{eqnarray}
M_{rem} &=& \dfrac{\sqrt{3 a'_1}}{8 \pi}M_P
\label{mr1}
\end{eqnarray}

Alternatively the value for remnant mass can also be obtained by minimising the entropy ({\ref{ms1}}),
\begin{equation}
\dfrac{dS}{dM}\;=\;0
\end{equation}
and looking at the second derivative $\left(\dfrac{d^2 S}{dM^2}>0\right)$. The result ({\ref{mr1}}) is reproduced.

 The most important fact is that the value ({\ref{mr1}}) is greater than the singular mass ({\ref{ms1}}) and also the critical mass ({\ref{mcr1}}). So we can say that the singularity will be avoided during the evaporation process and the reduced area will  be positive. At the same time we also managed to avoid any possibility of dealing with complex values for the thermodynamic entities $\left(since,~M_{rem}>M_{cr}\right)$.

The remnant value of area, reduced area, temperature and entropy are now expressed in terms of the coefficient $a'_1$. 
\begin{eqnarray}
A_{rem}&=&\dfrac{16\pi G^2M_{rem}^2}{c^4}\;=\;\dfrac{3a'_1}{64{\pi}^2}\dfrac{16\pi G^2M_P^2}{c^4}\\
A'_{rem}&=&\dfrac{a'_1L_P^2}{4\pi}\;=\;\dfrac{16\pi G^2M_{rem}^2}{3c^4}=\dfrac{A_{min}}{3}\\
T_{rem} &=& \dfrac{1}{\sqrt{a'_1}}\dfrac{M_Pc^2}{k_B} \\
S_{rem} &=& \dfrac{k_B}{16\pi} \left[ a'_1 - a'_1 \ln \left(a'_1 \right) \right] 
\end{eqnarray}

The expressions for different thermodynamic entities including the remnant mass involves only one parameter $a'_1$ as a measure of quantum gravity effect. If we put $a'_1=0$ we get back all our semi classical results and the remnant mass becomes zero. The final entropy and heat capacity become zero while the temperature becomes infinite.\\

\subsection{\bf Second order correction}
 The heat capacity with the second order contribution is given by
\begin{eqnarray}
C&=&\dfrac{k_B}{8\pi}\left[-\left(\dfrac{M_Pc^2}{k_BT}\right)^2+a'_1+3a'_2\left(\dfrac{k_BT}{M_Pc^2}\right)^2\right]
\label{c2}
\end{eqnarray}

The expression for remnant mass can be found either from zero heat capacity condition or by minimising the entropy. Adopting the first approach we will compute the remnant mass here. Replacing the value of $T^2$ from ({\ref{mt2}}) in ({\ref{c2}}) and equating the r.h.s. to zero we obtain the remnant mass 
\begin{equation}
\left(\dfrac{8\pi M_{rem}}{M_P}\right)^2 = \dfrac{1}{6a'_2}\left[-a'_1\left({a'_1}^2-13a'_2\right)+\left({a'_1}^2+5a'_2\right)\sqrt{{a'_1}^2+12a'_2}\;\right]
\end{equation}  

One can easily show that the r.h.s. is greater than $3a'_1$ which is the value ({\ref{mr1}}) for $\left(\dfrac{8\pi M_{rem}}{M_P}\right)^2$ with first order correction. The remnant mass is also greater than the critical mass ({\ref{mcr2}}) for positive $a'_1$ and $a'_2$. The difference between remnant mass and the critical mass for different values of $a'_1$ and $a'_2$ is shown in $fig.~4$ .

\section{\bf Discussion }
The laws of black hole thermodynamics are known to be modified by the presence of a generalised uncertainty principle (GUP) \cite{arz,vag,nuc,bin}.Here we have derived a new GUP (based on the presence of a minimal length scale $(L_P)$) which, at the lowest orders, is also shown to be compatible with string theory predictions. Using this GUP various aspects of Schwarzscild black hole thermodynamics were examined.

Our calculations were performed upto two orders (in $L_P$) of corrections. Corrected structures of mass-temperature relation, area theorem and heat capacity were obtained. The usual semiclassical expressions were easily derived.

An important consequence was that the black hole evaporation terminated at a finite mass. This (remnant) mass was found to be greater than either the critical mass (below which the thermodynamic variables become complex) or the singular mass (where the thermodynamic variables become infinite). Consequently the ill defined situations were bypassed. Also, contrary to standard results \cite{arz,vag} using GUP our modified area law ({\ref{sa1},\ref{sa2}}) is more transparent when expressed in terms of reduced area defined in ({\ref{a1}}).

To put our results in a proper perspective let us compare with earlier findings. A remnant mass was also found in \cite{adl} using stringy GUP and in \cite{xng} employing notions of tunnelling. In the first case the remnant mass was given by $M=M_P$ (which is consistent with our findings) and successfully avoided the singularity. However, in contrast to our analysis, the calculations were confined to the leading order only. Also, the remnant and the critical mass became identical. Hence it was not possible to distinguish between the termination of black hole evaporation and complexification of thermodynamic variables. The calculation of \cite{xng}, on the other hand, led to the result that although there was a remnant mass, the singularity problem could not be avoided.

\section*{\bf Acknowledgement}
One of the authors (S.G.) thanks the Council of Scientific and Industrial
Research (C.S.I.R), Government of India, for financial support. We also thank B. R. Majhi and S. K. Modak for discussions.
\begin{figure}[h]
\centering
\includegraphics[scale=.75]{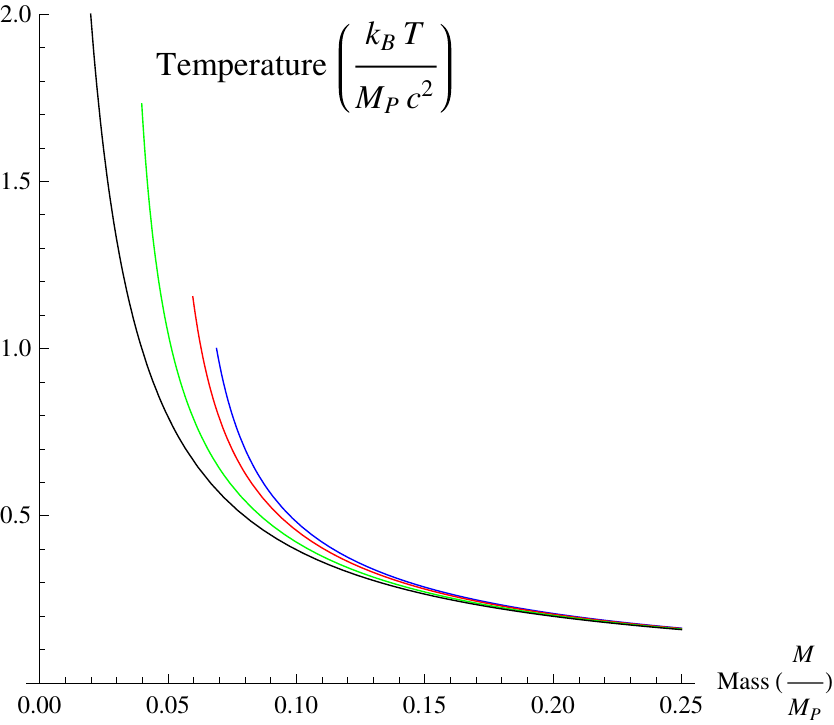}\\
\caption{ Temperature - mass curve with $a'_1 = 1$ (blue), $a'_1 = .75$ (red), $a'_1 = \dfrac{1}{3}$ (green) and $a'_1 = 0$ (black) [Semiclassical]$\vspace{.5cm} $ }

\includegraphics[scale=0.75]{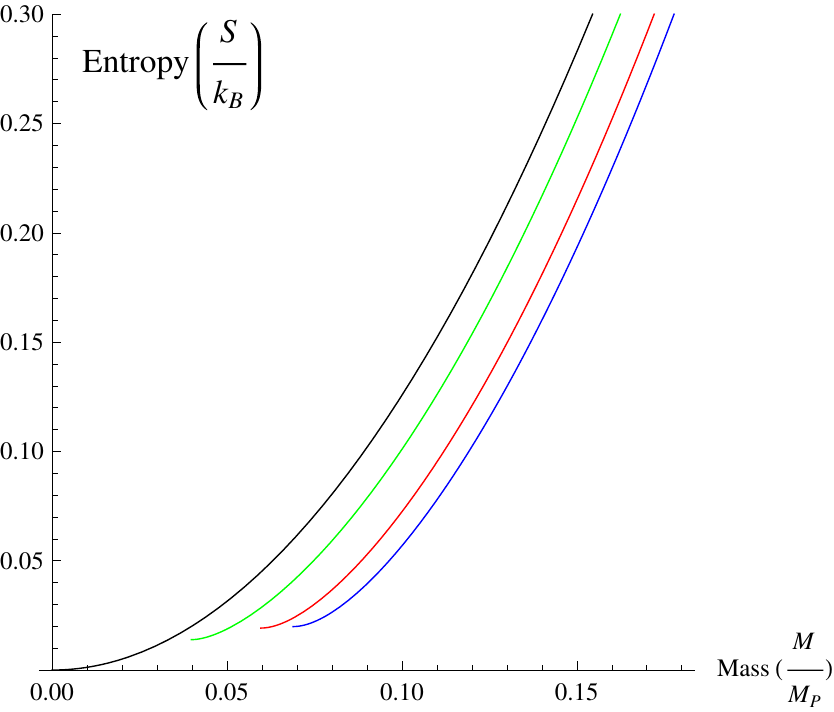}\\
\caption{ Entropy - mass curve with $a'_1 = 1$ (blue), $a'_1 = .75$ (red), $a'_1 = \dfrac{1}{3}$ (green) and $a'_1 = 0$ (black) [Semiclassical]$\vspace{.5cm} $ } 

\includegraphics[scale=0.75]{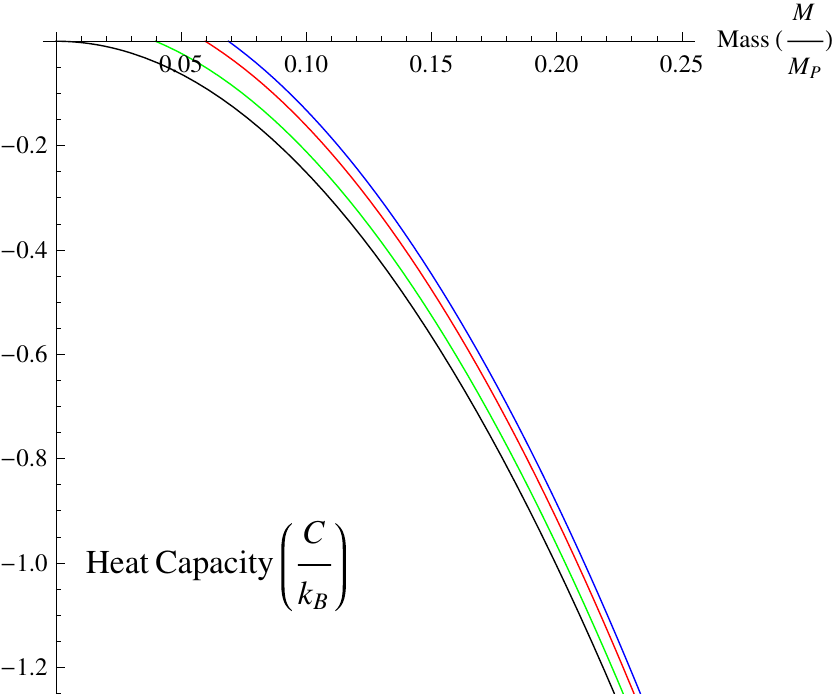}\\
\caption{ heat capacity - mass curve with $a'_1 = 1$ (blue), $a'_1 = .75$ (red), $a'_1 = \dfrac{1}{3}$ (green) and $a'_1 = 0$ (black) [Semiclassical] } 

\end{figure}

\begin{figure}[h]
\center
\includegraphics[scale=.7]{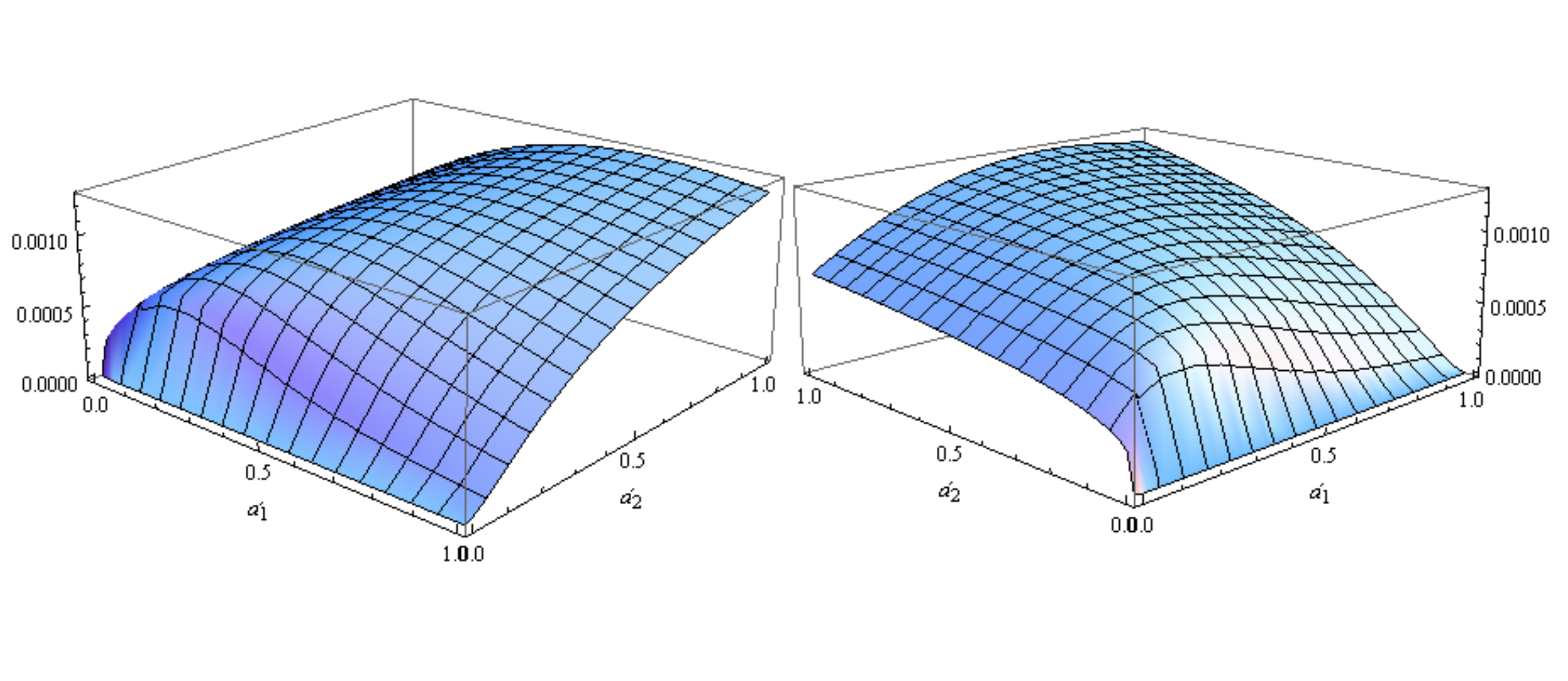}\\
\caption{ difference between the remnant mass and the critical mass $(M_{rem}-M_{cr})$  in unit of $M_P$ for different values of $a'_1$ and $a'_2$ shown from two different angles. The difference becomes zero only at $a'_2=0$. This point is not considered, since for $a'_2=0$, second order correction is not meaningful.}
\end{figure}
  


\end{document}